\documentclass{nature2}

\usepackage{lineno}

\usepackage{multicol}
\usepackage{amsmath}
\usepackage{booktabs}
\usepackage{isotope}
\usepackage{siunitx}
\usepackage{chemformula}
\usepackage{graphicx}
\usepackage{float}
\usepackage{footnote}
\makesavenoteenv{tabular}
\makesavenoteenv{table}
\usepackage{textgreek}
\newcommand{\EAunitless}{2.41578(7)} %EA without unit e.g. for Julias table
\newcommand{\EA}{\SI{\EAunitless}{eV}} %EA with units for text

\newcommand{\IP}{\SI{9.31751(8)}{eV}} %IP with units for text
\newcommand{\bandwidth}{\SI{12}{GHz}} %laser bandwidth in GHz
\newcommand{\electronegativity}{\SI{5.86665(7)}{eV}} %laser bandwidth in GHz
\title{The electron affinity of astatine}

\author{David Leimbach$^{1,2,3*}$, 
Julia Sundberg$^2$, 
Yangyang Guo$^4$, 
Rizwan Ahmed$^{5}$, 
Jochen Ballof$^{1,6}$, 
Lars Bengtsson$^2$, 
Ferran Boix Pamies$^1$, 
Anastasia Borschevsky$^4$, 
Katerina Chrysalidis$^{1,3}$, 
Ephraim Eliav$^{11}$, 
Dmitry Fedorov$^{7}$, 
Valentin Fedosseev$^1$, 
Oliver Forstner$^{8,9}$, 
Nicolas Galland$^{10}$,
Ronald Fernando Garcia Ruiz$^1$, 
Camilo Granados$^1$, 
Reinhard Heinke$^3$, 
Karl Johnston$^1$, 
Agota Koszorus$^1$, 
Ulli K\"oster$^{13}$, 
Moa K. Kristiansson$^{14}$, 
Yuan Liu$^{15}$, 
Bruce Marsh$^1$, 
Pavel Molkanov$^{7}$,
Luk\'{a}\v{s} F. Pa\v{s}teka$^{12}$, 
Joao Pedro Ramos$^1$, 
Eric Renault$^{10}$, 
Mikael Reponen$^{16}$,
Annie Ringvall-Moberg$^{1,2}$, 
Ralf Erik Rossel$^1$, 
Dominik Studer$^3$, 
Adam Vernon$^{17}$, 
Jessica Warbinek$^{2,3}$,
Jakob Welander$^2$,
Klaus Wendt$^3$, 
Shane Wilkins$^1$, 
Dag Hanstorp$^2$  
and Sebastian Rothe$^1$}

\begin{document}

\maketitle

\begin{affiliations}
 \item CERN, Geneva, Switzerland
 \item Department of Physics, University of Gothenburg, Gothenburg, Sweden
 \item Institut f\"ur Physik, Johannes Gutenberg-Universit\"at, Mainz, Germany
 \item Van Swinderen Institute for Particle Physics and Gravity, University of Groningen, Groningen, The Netherlands
   \item National Centre for Physics (NCP), Islamabad, Pakistan
  \item Institut f\"ur Kernchemie, Johannes Gutenberg-Universit\"at, Mainz, Germany
   \item Petersburg Nuclear Physics Institute - NRC KI, Gatchina, Russia
  \item Institut f\"ur Optik und Quantenelektronik, Friedrich-Schiller-Universit\"at Jena, Germany
 \item Helmholtz-Institut Jena, Jena, Germany
 \item CEISAM, Universit\'e de Nantes, CNRS, Nantes, France
 \item School of Chemistry, Tel Aviv University, Tel Aviv, Israel
 \item Department of Physical and Theoretical Chemistry \& Laboratory for Advanced Materials, Faculty of Natural Sciences, 
Comenius University, Bratislava, Slovakia
\item Institut Laue-Langevin, Grenoble, France
 \item Department of Physics, Stockholm University, Stockholm, Sweden
 \item Physics Division, Oak Ridge National Laboratory, Oak Ridge, Tennessee, USA
\item Department of Physics, University of Jyv\"askyl\"a, Jyv\"askyl\"a, Finland
\item School of Physics and Astronomy, The University of Manchester, Manchester, UK

\end{affiliations}

\begin{abstract}

One of the most important properties influencing the chemical behavior of an element is the energy released with the addition of an extra electron to the neutral atom, referred to as the electron affinity (EA). 
Among the remaining elements with unknown EA is astatine, the purely radioactive element 85.
Astatine is the heaviest naturally occurring halogen and its isotope $^{211}$At is remarkably well suited for targeted radionuclide therapy of cancer. 
With the At$^-$ anion being involved in many aspects of current astatine labelling protocols, the knowledge of the electron affinity of this element is of prime importance. In addition, the EA can be used to deduce other concepts such as the electronegativity, thereby further improving the understanding of astatine's chemistry.
Here, we report the first measurement of the EA for astatine to be \textbf{\EAunitless}~eV. This result is compared to state-of-the-art relativistic quantum mechanical calculations, which require incorporation of the electron-electron correlation effects on the highest possible level.
The developed technique of laser-photodetachment spectroscopy of radioisotopes opens the path for future EA measurements of other radioelements such as polonium, and eventually super-heavy elements, which are produced at a one-atom-at-a-time rate.

\end{abstract}

%%%%%%%%%%%%%%%%%%%%%%%%%%%%%%%%%%%%%%%% INTRODUCTION %%%%%%%%%%%%%%%%%%%%%%%%%%%%%%%%%%%%%%%%%%%%%
%\begin{multicols}{2}
\section*{Introduction}
\begin{figure*}[h]
    \centering
    \includegraphics[trim= 0mm 0mm 0mm 0mm , clip, width=0.7\linewidth]{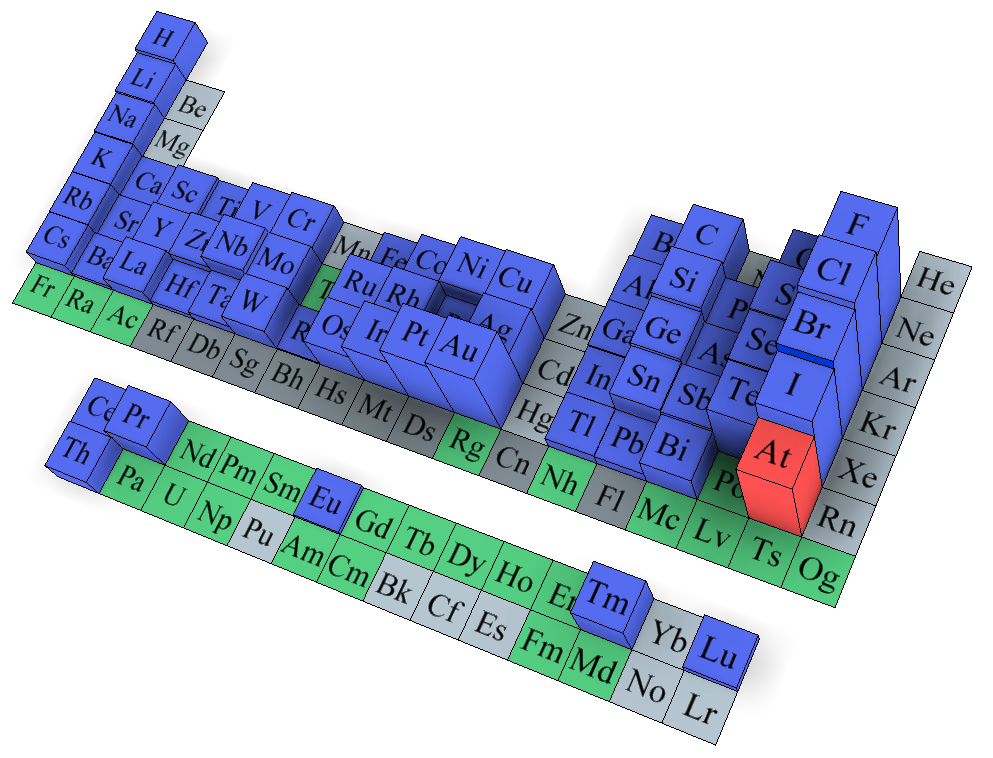}
        \caption{Electron affinities across the periodic table. The height corresponds to the measured value of the electron affinity of the corresponding element\cite{AndHauHot99,Thorium}. Astatine is highlighted in red. Blue indicates elements that are experimentally determined to have a positive EA, i.e. to form stable negative ions. Elements that are predicted to form stable negative ions but not yet have been experimentally investigated are indicated in green, while those in light grey are predicted to not form a stable negative ion, i.e. have an negative EA.
Finally, elements that neither have been experimentally observed nor investigated theoretically, are  indicated with dark grey.}
    \label{fig: PT}
\end{figure*}

Chemistry is all about molecule formation through the creation or destruction of chemical bonds between atoms and relies on an in-depth understanding of the stability and properties of these molecules. 
Most of these properties can be traced back to the molecule's constituents, the atoms.
Thus, the intrinsic characteristics of chemical elements are of crucial importance in the formation of chemical bonds.
%, and the value of the electron affinity (EA) is paramount.\\ 
The electron affinity (EA), one of the most fundamental atomic properties, is defined as the amount of energy released when an electron is added to a neutral atom in the gas phase. %\cite{Goldbook}. 
Large EA values characterize electronegative atoms, i.e. atoms that tend to attract shared electrons in chemical bonds.
Hence, the EA informs about the subtle mechanisms in bond making between atoms, and it also reveals information about molecular properties such as the dipole moment or the molecular stability. 
Since the attraction from the nucleus is efficiently screened by the core electrons, the value of the EA is mainly determined by electron-electron correlation.
Hence, negative ions are excellent systems to benchmark theoretical predictions that go beyond the independent particle model. \\
The EA also enters into the definition of several concepts, notably the chemical potential within the purview of conceptual density functional theory (DFT), promoted by Robert G. Parr\cite{Parr1978}, and the chemical hardness which is the core of the hard and soft acids and bases (HSAB) theory, introduced by Ralph G. Pearson in the early 1960s\cite{Pearson1963}.
Robert S. Mulliken used the EA in combination with the ionization energy (IE), the minimum amount of energy required to remove an electron from an isolated neutral gaseous atom, to develop a scale for quantifying the electronegativity of the elements\cite{Mulliken1934}. 
The usefulness of these concepts for chemists, especially in the field of reactivity, has been amply demonstrated in recent decades\cite{Geerlings2003, Chattaraj2006}. \\
The atomic IEs, which essentially are determined by the Coulomb attraction between the electrons and the nucleus, show a specific and well understood variation along the periodic table of elements. Starting from lowest values in the lower left corner at the heaviest alkalines, a mostly steady trend towards higher values is observed both towards ligther elements with similar chemical behaviour in one column and along rows to the right side of the chart with halogenes and noble gases, with only few exceptions. 
Conversely, the EAs display comparably strong irregularities and variations across the periodic table, as shown in Fig. \ref{fig: PT}. 
A number of elements such as all the noble gases do not form stable negative ions at all, and thus have negative EAs.
The group of elements with the largest EAs are the halogens. 
As in most other groups of elements, no monotonic trend is observed here when progressing along the rows of the periodic table, with chlorine exhibiting the largest EA (\SI{3.612 725(28)}{eV}) of all elements\cite{AndHauHot99, Thorium}.\\
The EA of the heaviest naturally occurring elements in the halogen group,  astatine, has not been measured to date. Indeed, little is known of the chemistry of this rare element: it is not only one of the rarest of all naturally occurring elements\cite{asimov}, but the minute amounts that can be produced artificially prevent the use of conventional spectroscopic tools. 
For instance, while astatine was discovered in the 1940s\cite{Corson1940,Thornton2019}, it is only recently that the IE of astatine was measured through a sophisticated on-line laser-ionization spectroscopy experiment at CERN-ISOLDE\cite{Rothe_2013}.\\
%(no halogen-specific characteristics have been reported so far for the recently discovered tennessine). 
However, the EA(At) has been predicted with various quantum mechanical methods\cite{SiFis18, FinPet19, Mitin2006Two-componentMethods, LiZhaAnd12, Borschevsky2015IonizationAt, Sergentu2016, ChaLiDon10}.  
Hence, an experimental determination of EA(At) is of fundamental interest, both to test sophisticated atomic theories and to gain precise knowledge about the chemical properties of this element. 
The measurement of the EA(At) is also of practical interest regarding the envisaged medical applications of astatine, since its chemical compounds are currently studied for use in cancer treatment:
$^{211}$At, available in nanogram quantities only through synthetic production methods, is a most promising candidate for radiopharmaceutical applications via targeted alpha therapy (TAT)\cite{Zalutsky2011,MulfordTAT, teze:in2p3-01529705}, due to its favorable half-life of about \SI{7.2}{h} and its cumulative $\alpha$-particle emission yield of \SI{100}{\%}. However, in order to successfully develop efficient radiopharmaceuticals, a  better understanding of astatine's basic chemical properties is required\cite{Wilbur2013}.\\
The interest in the experimental determination of the EA notably lies in current labelling protocols that aim at binding astatine to tumor-targeting biomolecules: in many cases, the chemical reactions involve an aqueous astatine solution in which the astatide anion (At$^-$) readily forms.
In addition, a current problem for the investigated $^{211}$At-radiopharmaceuticals is the significant \textit{in vivo} de-labelling, releasing At$^-$ that could damage healthy tissues and organs of the patient\cite{teze:in2p3-01529705, Vaidyanathan2008, Wilbur2008}. 
In order to describe these reaction kinetics as well as the stability of astatine compounds, knowledge of the electron binding energy of the atomic  anion, i.e. the EA, is required. \\
In this paper, we present the first experimental determination of the electron affinity of astatine. 
The measured value is then compared to independent results from state-of-the-art relativistic quantum mechanical calculations carried out alongside the measurement.

%%%%%%%%%%%%%%%%%%%%%%%%%%%%%%%%% RESULTS %%%%%%%%%%%%%%%%%%%%%%%%%%%%%%%%%
\section*{Results}

\subsection{Laser photodetachment of astatine.}
\begin{figure*} [h]
    \centering
    \includegraphics[trim= 0mm 0mm 0mm 38mm , clip, width=0.8\linewidth]{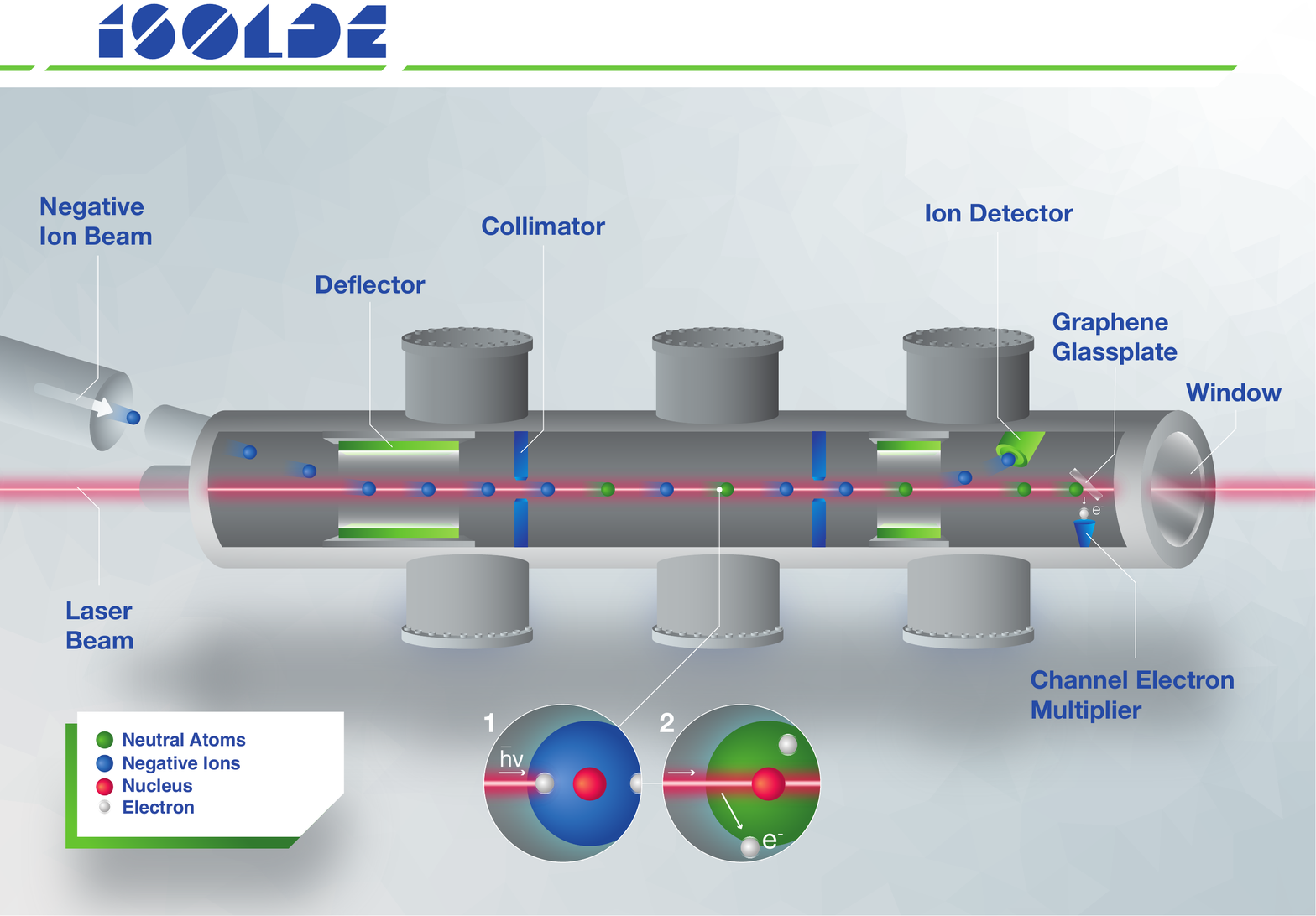}
        \caption{Schematic diagram of the experimental setup. From left to right: A beam of negative astatine ions is guided into the Gothenburg ANion Detector for Affinity measurements by Laser PHotodetachment (GANDALPH)\cite{iodine128,Leimbach_EMIS}, where the ion beam is overlapped with a frequency tuneable laser beam in the interaction region in either co- or counter-propagating geometry. By absorbing a photon (Inset 1), an electron can gain enough energy to be ejected from the ion, thereby creating a neutral atom (Inset 2). After the interaction region, the charged particles are deflected into an ion detector, while neutralized atoms continue moving straight to the graphene-coated glass plate downstream and create secondary electrons, which are detected by a channel electron multiplier\cite{Warbinek}.}
    \label{fig:GANDALPH}
\end{figure*}

Due to its scarcity and short half-life, artificial production of astatine is required to perform any experiment on this element. 
Thus, a laser photodetachment threshold spectrometer was coupled to an on-line isotope separator at the CERN-ISOLDE radioactive ion beam facility\cite{Catherall_2017}.
Here, At$^-$ ions were produced through nuclear spallation reactions of thorium nuclei, induced by a bombardement of highly energetic proton projectiles and subsequently ionized in a negative surface ion source coupled to a mass separator (see Fig. \ref{fig:RIB} in the Methods section).
A negative ion beam of $^{211}$At was extracted and superimposed with a laser beam in the GANDALPH spectrometer (Fig. \ref{fig:GANDALPH}).  
The yield of neutral atoms produced in the photodetachment process, ${\mathrm{At}^- + \mathrm{h}\nu}  \rightarrow \mathrm{At} + \mathrm{e}^-$, was recorded as a function of the photon energy $\mathrm{h}\nu$, where $\nu$ is the laser frequency and $\mathrm{h}$ is Planck's constant.

The general behavior of the photodetachment cross section $\sigma$ just above the threshold is described by Wigner's law\cite{Wigner1948OnThresholds}: $\sigma= a+b\cdot E^{l+1/2}$, where $a$ is the background level, $b$ the strength of the photodetachment process, $l$ the orbital angular momentum quantum number of the outgoing electron, $E=E_\mathrm{photon}-\mathrm{EA}$ is the energy of the ejected electron and $E_\mathrm{photon}=\mathrm{h}\nu$ the photon energy.\\
In At$^-$, the electron is detached from a $p$-state. Close to the threshold, the angular momentum of the outgoing electron will then be  $l=0$ due to the selection rules ($\Delta l= \pm 1$) and the centrifugal barrier preventing the emission of a $d$-wave electron ($l=2$)\cite{Pegg_2004}.
The \ch{At-} ion is a closed shell system with no internal structure.  The ground state $6p^{5}~^{2}P_{3/2}$ of the $^{211}$At atom, on the other hand, with a total angular momentum of $J={3/2}$  and nuclear spin $I=9/2$, is split into four hyperfine levels. This splitting was recently measured with high precision by Cubiss \textit{et al.}\cite{Cubiss_2018}.  
The relative strengths of these four photodetachment channels are given by the multiplicity of the final hyperfine structure levels, i.e. $2F+1$, where $F=I+J$ is the total angular momentum of the atom, spanning from $|I - J|$ to $|I + J|$, i.e. 3,4,5,6\cite{HanGus92}.
The energy dependence of the cross section for photodetachment of astatine near the threshold can be described by the function 
\begin{linenomath}\begin{equation}
\sigma(E_\mathrm{photon}) = a + b \sum_{F = 3}^6 (2F+1) \sqrt{E_\mathrm{photon} - (\mathrm{EA} + \mathrm{E_{hfs,F}})}~\Theta\mathopen{}\left(E_\mathrm{photon}- (\mathrm{EA} + \mathrm{E_{hfs,F}})\right)
\label{eq:cross_section2}
\end{equation}\end{linenomath}
%\begin{equation}
%\sigma = 
%\begin{cases}
%a + b\sum\limits_{F=3}^{6} 
%(2F +1) (h\nu-(EA+E_{\text{hfs}}))^{1/2} &\text{for } $h\nu$ \geq EA \\
%a &\text{for } $h\nu < EA$,
%\end{cases}
%\label{eq:cross_section2}
%\end{equation}
where $\Theta\mathopen{}\left(E- (\mathrm{EA} + \mathrm{E_{hfs,F}})\right)$ is a heaviside function and $\mathrm{E_{hfs,F}}$ is the energy of the hyperfine levels of the $^{211}$At atomic ground state, differing by less than \SI{23}{\micro eV} between the contributing levels.\\
\begin{figure*} [h]
    \centering
    \includegraphics[trim= 22mm 10mm 20mm 20mm , clip, scale=0.6]{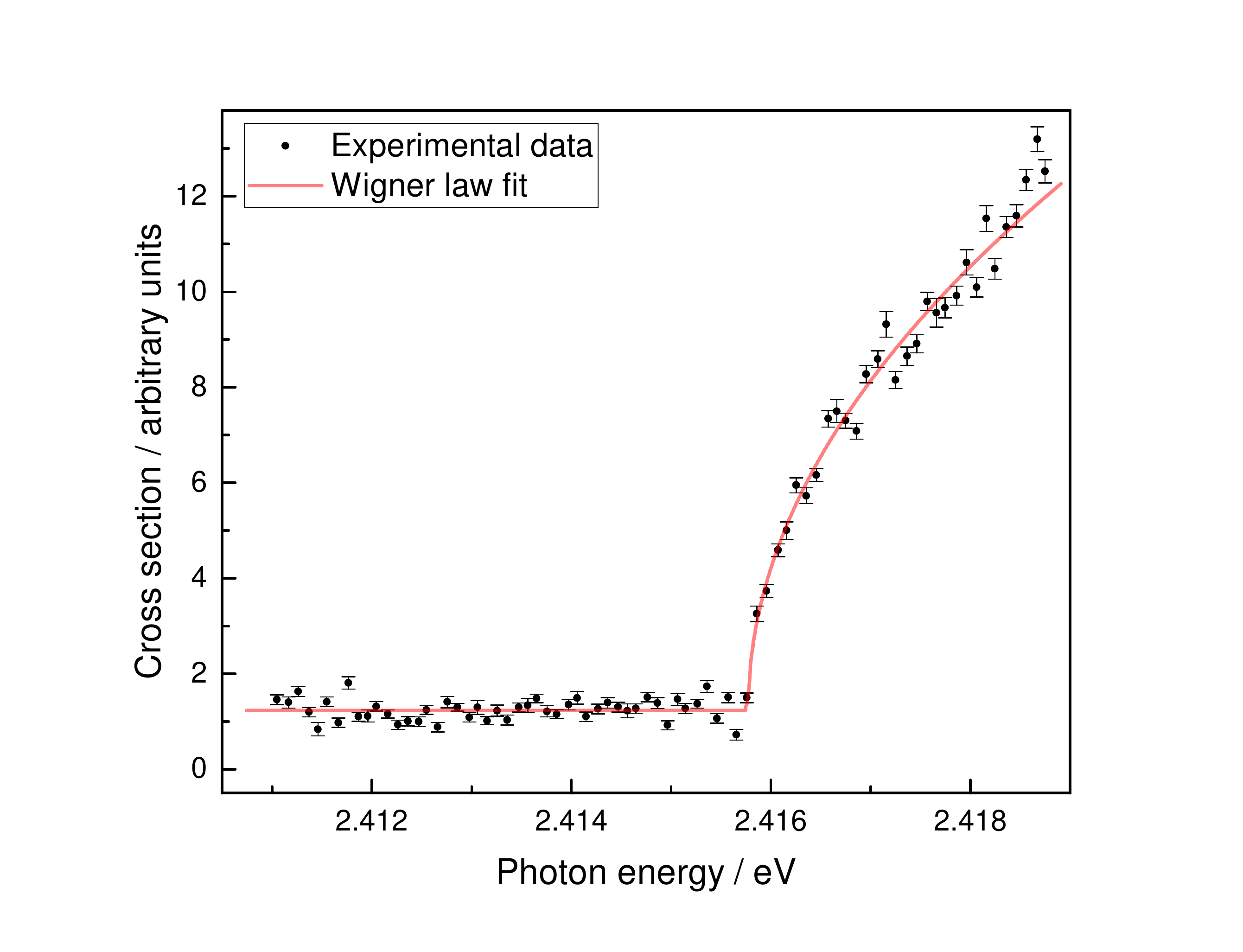}
    \caption{Threshold scan of the photodetachment of astatine. Neutralization cross section is measured as a function of the photon energy. The data points are the experimental measurements with one standard error represented with error bars, and the solid line is a fit of Eq. \ref{eq:cross_section2}. The onset corresponds to the EA of $^{211}$At.}
    \label{fig:Unshifted_EA}
\end{figure*}
The photon energy (i.e. laser frequency) was scanned from below the threshold to well above all four hyperfine levels in the ground state of $^{211}$At.
In total, six threshold scans were performed with laser and ion beam co- and counter-propagating, respectively.
Fig. \ref{fig:Unshifted_EA} shows the measured neutralization cross section $\sigma(E_\mathrm{photon})$ as a function of the photon energy, corrected for the Doppler shift, for the sum of all threshold scans with co-propagating ion and laser beams.\\
The statistical error of the measurement is dominated by the laser bandwidth of \bandwidth,  corresponding to \SI{50}{\micro eV}. The contribution to the statistical uncertainty from all other effects are smaller than \SI{0.1}{\micro eV}, as discussed further in the Methods section, and can hence be neglected.
Systematic errors arise due to instabilities of the ion beam energy and the determination of the photon energy. 
The combined systematic error of photon energy and beam energy is estimated to be smaller than \SI{20}{\micro eV} by comparing two reference 
measurements of stable $^{127}$I which were performed before and after the experiment on astatine, under the same experimental conditions. \\
Including both systematic and statistical errors, the resulting value of EA(At), calculated by the geometric mean of the photodetachment thresholds measured in the co- and counter-propagating geometries, was determined to be \EA.

\subsection{Theoretical Calculation.}

Alongside the measurements, state-of-the-art calculations of the electron affinities of astatine and of its lighter homologue, iodine ($^{127}$I) were carried out. 
The results for EA(I) served to assess the performance and the expected accuracy of the computational method.
The calculations were carried out with the DIRAC15 program package\cite{DIRAC15} using the single reference coupled cluster approach in the framework of the Dirac-Coulomb Hamiltonian (DC-CCSD(T)), which is considered to be extremely powerful for treatment of heavy many-electron systems. Large, saturated basis sets\cite{Dya06} were used in these calculations, and extrapolation to the complete basis set limit was performed. 
The correction from perturbative to the full triple excitations, +$\Delta$T, and the contribution of the perturbative quadruple excitations, +(Q), were evaluated\cite{PasEliBor17}. To further improve the precision we have also accounted for the Breit interaction and the quantum electrodynamics (QED) contributions; the latter were calculated using the model Lamb shift operator (MLSO) of Shabaev \textit{et al.}\cite{ShaTupYer15}. Further computational details can be found in the Methods section. The contributions of higher order excitations and Breit and QED corrections are added to the DC-CCSD(T) EAs to obtain the final values. 
\begin{table*} [h]
\caption{Comparison of computational results  including higher order contributions and experimentally determined values of the electron affinities of iodine and astatine.}
\label{table1}
\centering
%\footnotesize

\begin{tabular}{ l| c| c } 
 \hline\hline
 Method & EA(I)/eV & EA(At)/eV\\ 
 \hline
 DC-CCSD(T) &3.040&  2.401 \\ 
%  $\Delta$T & 0.079 &0.092\\ 
    +$\Delta$T(Q)& 0.008& 0.007\\
 +Breit & 0.003 &0.003\\ 
   +QED & 0.003 &0.003\\ 
   \textbf{Final theor.} &~~~~~ \textbf{3.055(16)}&~~~~~~~\textbf{2.414(16)} \\ \hline
 Exp. & ~~~~~~~~~~~~~~~~~~3.059 0463(38)\cite{Pel_ez_2009} &~~~~~~~~ \textbf{\EAunitless} \\ 
 \hline\hline
\end{tabular}

\end{table*}
The computational scheme outlined above was previously applied to the determination of the IE and the EA of gold, yielding \SI{}{meV} accuracy\cite{PasEliBor17}.
Using our knowledge of the magnitude of the various effects, we are able to set a conservative uncertainty of \SI{\pm0.016}{eV} on the computed values (see Methods section for further details). Hence, the expected value of the EA(At) from the theoretical calculations is \SI{2.414(16)}{eV}.
The results for iodine and astatine, including the break-down of the various higher order contributions are presented in Tab. \ref{table1} and compared to the experimental value. The final result of the electron affinity calculation for iodine lies within \SI{0.004}{eV} of the measured value of \SI{3.059 0463(38)}{eV}\cite{Pel_ez_2009}.

%%%%%%%%%%%%%%%%%%%%%%% DISCUSSION  %%%%%%%%%%%%%%%%%%%%%%%%%%%%%%%%%%%%%%%%

\section*{Discussion}
Over the years, many attempts were made to calculate the EA of astatine. However, the high atomic number and thus the need of refined treatments of relativity as well as the dominance of the electron correlation effects made this a challenging task.
With the given uncertainties, our computed value is in excellent agreement with the experiment.
This clearly demonstrates that careful, systematic, and as complete as possible inclusion of higher-order correlation and relativistic contributions is necessary for achieving benchmark accuracy.
Hence, our measured EA(At) represents a sharp test for assessing theoretical methods used to study the chemistry of heavy and super-heavy elements.
For a more detailed comparison of our computed results with previous theoretical investigations we refer the reader to the Methods section.\\
%A striking application of the EA(At) is the investigation of astatine's chemistry which heavily relies on molecular modeling due to its limited supply and radioactive nature.\\
Our result of the EA of astatine, \EA, indicates that among the halogen elements, astatine has the lowest EA. On the other hand, its EA remains larger than the measured values of all elements in other groups of the periodic table. 
Therefore, this value is consistent with the propensity of halogens to complete their valence shell on gaining one extra electron. 
%With astatine, the fitfth place on a scale as crucial for chemistry as the electronic affinity has been filled, where only nine elements exhibit higher EAs than \SI{2}{eV}. 
For the halogen elements, the significance of large EAs is the strong tendency to form anions in aqueous solution. In fact, the redox potential associated with the formation of At$^-$ is primarily determined by its EA, and to a lesser extent by the difference of Gibbs' free energy of solvation between the anion and the neutral atom. In addition to the EA, the IE contributes also to the determination of the nature of elemental forms of astatine in aqueous solutions:
the Pourbaix (potential/pH) diagram of astatine shows coexistence of the \ch{At+} and \ch{At-} ions, whose dominance domains are governed by the redox potential E$^\circ$(\ch{At-}/\ch{At+}), which directly depends on the EA(At) and IE(At)\cite{Champion2010,Champion2011}.\\
The usefulness of the EA for a better understanding of astatine's chemistry is also shown through the deduction of the electronegativity, softness, hardness, and the electrophilicity index, which are shown in Tab. \ref{tab:theory_comp}, together with the respective definitions.
The electronegativity of astatine is determined to be $\chi _M= $\SI{5.87}{eV} according to the Mulliken scale, which is  significantly lower than that of hydrogen ($\chi _M=$ \SI{7.18}{eV}), supporting the calculated bond polarization towards the hydrogen atom in the \ch{H-At} molecule\cite{Pilm2014,SAUE1996}. Hence, it must be named hydride instead of hydrogen halide as opposed to all other halogen-hydrogen molecules, where the halogen is usually the negatively charged atom. 
\begin{table*}
 \caption{Values and definitions of properties of astatine derived from the EA and IE.}
    \label{tab:theory_comp}
    \centering
    \begin{tabular}{lll}
    %\toprule 
    \hline \hline
    Property & Definition & Value \\
    \midrule
    Electron affinity & $EA$ & \EA  \\ \vspace{0.12cm}
    Ionization energy & $IE$  & \IP \cite{Rothe_2013}  \\  \vspace{0.12cm}
    Electronegativity &  $\chi_M =\frac{IE+EA}{2}$ & \electronegativity  \\ \vspace{0.12cm}
    Hardness & $\eta =\frac{IE-EA}{2}$ & \SI{3.45087(7)}{eV} \\ \vspace{0.12cm}
    Softness & $S=\frac{1}{2\eta}$ & \SI{0.14489(2)}{eV^{-1}} \\ \vspace{0.12cm}
    Electrophilicity & $\omega = \frac{\chi^2_M}{2 \eta}$  & \SI{4.98680(16)}{eV}  \\
    \hline \hline\\
    
     \end{tabular}
\end{table*}
Additionally, the intermediate value of $\chi _M$(At) between the electronegativities reported for boron (\SI{4.29}{eV}) and carbon (\SI{6.27}{eV}) atoms, allows us to anticipate different polarizations for At-B and At-C bonds. This simple analysis is of high relevance for the use of astatine in nuclear medicine.
The applications in targeted alpha therapy are currently hindered by the rapid de-astatination of carrier-targeting agents that occurs \textit{in vivo}.
In radiosynthetic protocols\cite{Vaidyanathan2008,Wilbur2008}, most reported biomolecules of interest have been labeled with $^{211}$At by formation of At-C or At-B bonds.
The greater stability observed \textit{in vivo} for the At-B bonds could be related to the polarization of those bonds towards the astatine atom\cite{AYED2016156}. The electrophilicity index is particularly relevant in view of the currently prevalent approach for the $^{211}$At-radiolabelling, which is to bind astatine to carrier molecules through an electrophilic substitution\cite{Vaidyanathan2008,Wilbur2008}. In addition, recent studies have illustrated how the electrophilicity of the astatine atom modulates the ability of astatinated compounds to form stabilizing molecular interactions known as halogen bonds\cite{Graton2018,GaMontavon2018}.
%The same trends can be drawn from the scale introduced very recently by Rahms\cite{Rahm2019}, which provides electronegativities for elements 1 to 96. However, the electronegativity of At in that study was derived from quantum chemical calculations, while the present value relies on measured properties. 
The moderate value of hardness, $\eta$(At) = \SI{3.45}{eV}, is consistent with the observed high affinity of astatine in direct attachment experiments with proteins bearing soft sulfur donor groups\cite{VISSER1981905}, according to the hard and soft (Lewis) acids and bases (HSAB) theory ($\eta$(S) = \SI{4.14}{eV} for the S atom\cite{Pearson1988}).\\
%Indeed, soft acids react faster and form stronger bonds with soft bases according to the HSAB theory, and 211At forms for instance relatively stable interactions with SH containing proteins while a weaker binding is observed for proteins lacking a free SH-group
The list of chemical descriptors presented in Tab. \ref{tab:theory_comp} represents a significant advance over the computed data reported by Paul Geerlings and co-workers\cite{Giju2005}, and may be regarded as basic properties which will serve as the foundation for the design and the assessment of innovative astatine radiopharmaceuticals by theoretical and experimental chemists.

%%%%%%%%%%%%%%%%%%%%%%%%%%%%%%%%%%%%%%%%%%%%%%%%%%%%%%%%%%%%%%%%%%%% CONCLUSION  %%%%%%%%%%%%%%%%%%%%%%%%%%%%%%%%%%%%%%%%%%%%%%%%%%%%%%%%%%%

\section*{Conclusion}
We have carried out the first measurement of the electron affinity of astatine and determined it to be EA(At)$=$\EA.
In addition, relativistic calculations carried out alongside the experiment are in excellent agreement with the experimental results, supporting the reliability and accuracy of both the experimental technique and the theoretical description. 
The EA of astatine is thus an excellent case for benchmarking theoretical models in atomic physics since it requires a full relativistic many-body treatment that also includes Breit and QED effects. These theoretical models can then be applied to the chemistry of elements heavier than astatine.  \\
By combining the present result with the recent measurement of the ionization energy of astatine\cite{Rothe_2013}, we were able to determine several fundamental chemical properties of this element: namely the electronegativity, softness, hardness and electrophilicity.
For instance, it can be concluded from our results, that in the astatine-hydrogen molecule, in contrary to other hydrogen halides, the hydrogen atom is more electro-negative than the halide. Hence, according to chemical nomenclature this molecule should be called astatine hydride rather than hydrogen astatide.
\\
As $^{211}$At is a promising candidate for targeted alpha therapy, these properties have direct implications for its use in cancer treatments. Most of $^{211}$At-radiopharmaceuticals suffer from \textit{in vivo} release of astatide (At$^{-}$) 
and the development of radiosynthetic procedures so far is severely hampered by the limited knowledge of the chemical properties of this element. Hence, the new information about astatine's chemical properties presented here will be of great importance in the development of innovative radio-labelling protocols.\\
Finally, the on-line technique presented in this work enables further EA measurements of artificially produced, short-lived radioactive elements with high precision. Furthermore, our theoretical methods were demonstrated to be capable of accurately treating heavy elements with a high number of electrons, paving the way for both experimental and theoretical studies of superheavy elements. 

%%%%%%%%%%%%%%%%%%%%%%%%%%%%%%%%% METHODS %%%%%%%%%%%%%%%%%%%%%%%%%%%%%%%%%

\begin{methods}

%%%%%% Astatine ion production %%%%%%%%%%%%%%

\subsection{Negative astatine ions.}
Astatine isotopes were produced at the CERN-ISOLDE radioactive ion beam facility\cite{Catherall_2017}. A proton beam with an energy of \SI{1.4}{GeV} provided by the CERN accelerator complex impinged onto a thick Th/Ta mixed foil target, which was resistively heated to \SI{1450}{^\circ C}. A schematic view of this process is given in Fig. \ref{fig:RIB}.
The reaction products diffused from the target matrix and effused into an ISOLDE-MK4 negative surface ion source\cite{VOSICKI1981307}, comprised of a hot tantalum transfer tube and a \ch{LaB6} surface ionizer pellet heated to \SI{1300}{^\circ C}.
\begin{figure*} [h]
    \centering
    \includegraphics[trim= 0mm 0mm 0mm 0mm , clip, width=0.8\linewidth]{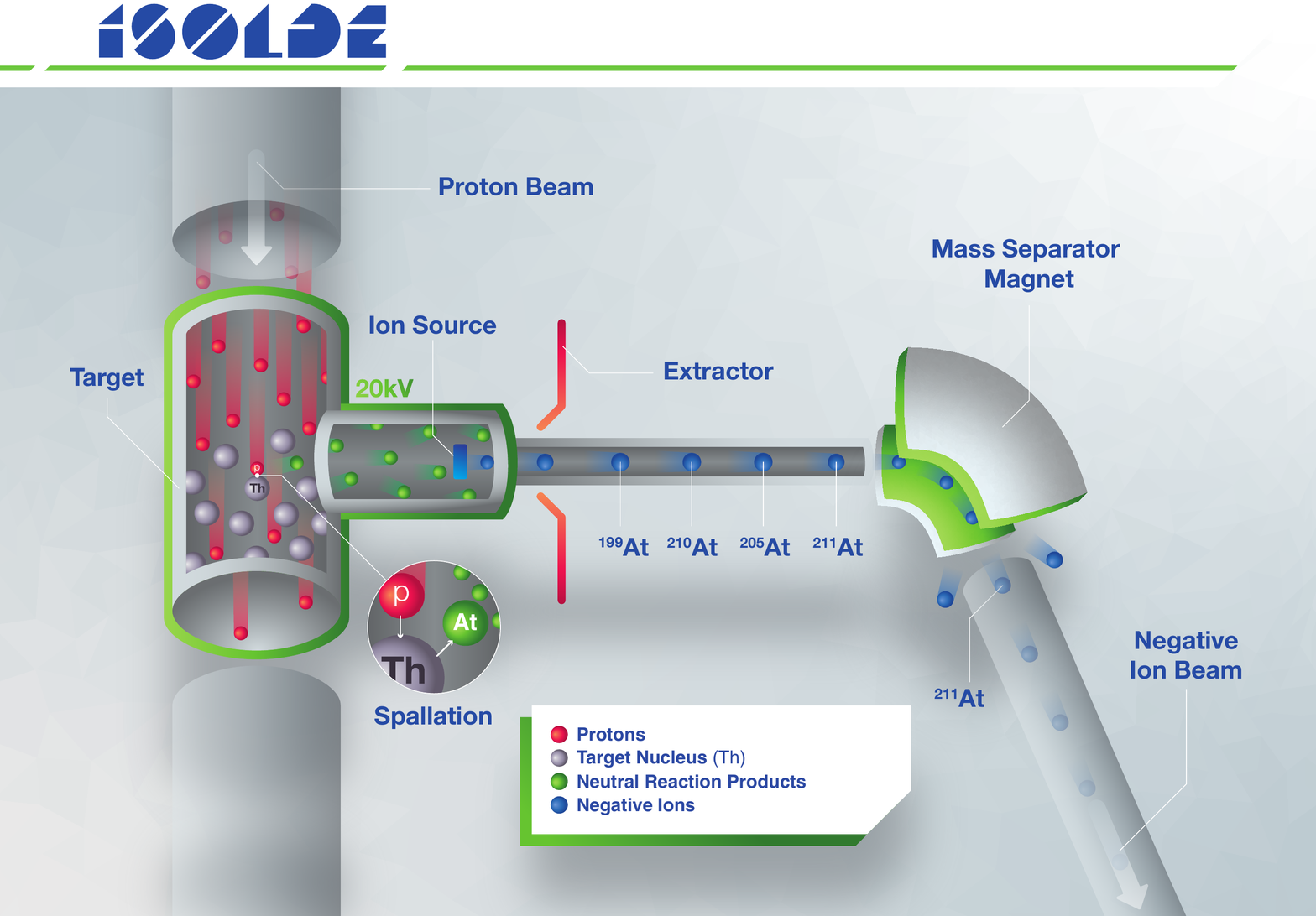}
        \caption{Production of a negative astatine ion beam. Astatine atoms are created in a spallation reaction of thorium with \SI{1.4}{GeV} protons. Subsequently, the atoms are negatively ionized and extracted as a mono-energetic beam with an energy of \SI{20}{keV}. The $^{211}$At isotopes are then mass separated with an electromagnetic mass separator and directed to the GANDALPH spectrometer.}
    \label{fig:RIB}
\end{figure*}
Thermionic electrons emitted from the hot \ch{LaB6} surface were deflected with a \SI{0.04}{T} permanent magnetic field and absorbed in a dedicated electron collector. Negative ions produced on the hot surface were accelerated across a \SI{20}{kV} extraction potential and thereafter directed through the ISOLDE general purpose mass separator magnet (GPS). The resolution of the mass separator was sufficient to select a single isobar, which in our case was $^{211}$At. \\
In order to ensure stable astatine beam intensity throughout the experiments, the pulsed proton impact on the target was  distributed equidistant in time with an average current of about \SI{1.8}{\micro A}. An average ion current of about \SI{600}{fA} (\SI{3.75E6}{particles/s}) of $^{211}$At$^-$  was measured using a Faraday cup (FC) inserted in the beam path just before the experimental chamber.

%%%%%%% Laser system %%%%%%%%%%%%%%%%%%%%%%
\subsection{Laser setup.}
The phototodetachment experiment was performed using a part of the ISOLDE RILIS (Resonance Ionization Laser Ion Source) laser system which normally serves for production of positively charged ion beams\cite{Fedosseev_2017}. 
In particular, laser radiation tunable in the range of \SI{2.384}{eV} to \SI{2.53}{eV} (\SI{490}{nm} to \SI{520}{nm}) was generated by a commercial dye laser (\textit{Sirah Laser- und Plasmatechnik GmbH Credo Dye}) operated with an ethanol solution of Coumarin 503 dye. 
This laser was pumped by the third harmonic output (\SI{3.4925}{eV}) of a pulsed Nd:YAG INNOSLAB laser (\textit{EdgeWave GmbH, model CX16III-OE}) with a \SI{10}{kHz} pulse repetition rate. 
Beam delivering optics comprising a set of lenses and mirrors were installed to transport the dye laser beam from the RILIS laboratory to the GANDALPH photodetachment apparatus over a distance of about \SI{15}{m}.
In the laser-ion beam interaction region, the laser power was in the range of 20-\SI{30}{mW}.
Typical values of the spectral bandwidth and pulse duration emitted by the dye laser were \SI{12}{GHz} and \SI{7}{ns}, respectively.
The laser radiation frequency was scanned in the range of \SI{2.411}-\SI{2.4301}{eV} (\SI{510}{nm}-\SI{514}{nm}), determined according to earlier theoretical predictions of the EA(At)\cite{Borschevsky2015IonizationAt}.
The photon energy of the laser radiation was measured continuously using a wavelength meter (WS7 model from HighFinesse/\AA ngstrom).

%%%%%% Laser photodetachment 
\subsection{Collinear laser photodetachment threshold spectroscopy with GANDALPH.}
The Gothenburg ANion Detector for Affinity measurements by Laser PHotodetachment (GANDALPH), illustrated in Fig. \ref{fig:GANDALPH}, is a detector designed for measurements of the EA of radioactive elements by collinear laser photodetachment\cite{iodine128, Leimbach_EMIS}.
Electrostatic beam steering and ion optical elements are used to superimpose a continuous negative ion beam with a pulsed laser beam within the interaction region of the GANDALPH spectrometer, which is defined by two apertures of \SI{6.0}{mm} diameter placed \SI{500}{mm} apart.
The experimental layout allows both co- and counter-propagating geometries for laser and ion beams.\\
When a negative ion absorbs a photon of sufficient energy, its extra electron can be detached, creating a fast moving neutral atom. The Doppler shift resulting from the velocity of the ion beam in reference to the detector and laser rest frame, can be eliminated to all orders by taking the geometrical mean of the measurements which are recorded in co-  and counter-propagating geometry of the laser and the ion beam, respectively. \\
Subsequent to the interaction region, all charged particles are deflected into either a FC or a channel electron multiplier (CEM), allowing for continuous monitoring of the ion beam intensity. Neutral atoms proceed forward and impinge on a target made of a graphene coated quartz plate\cite{Warbinek, iodine128 ,Hanstorp_1992}. \\
Secondary electrons created by the impact of the neutral atoms on the target are extracted and deflected into a second CEM (\textit{DeTech Channeltron XP-2334}), placed off-axis and biased with a potential of \SI{2.2}{kV}. 
The signal originating from the CEM is amplified with a fast pulse amplifier (\textit{FAST TA2000B-2}) by a factor of 40 and fed into a gated photon counter (\textit{Stanford Research Systems SR400}) connected to a computer.
A data acquisition cycle is triggered by the signal of the photoelectrons resulting from the laser pulse impinging on the glass plate target. Due to the time of flight from the interaction region to the glass plate, the neutral atoms created in the photodetachment proccess arrive in the time window \SI{2.2}-\SI{4.9}{\micro s} after the photon impact. Hence, the data acquisition is set to record the signal within this time window after the trigger.
Background measurements are performed simultaneously by setting a second measurement gate of the same width but delayed by \SI{12}{\micro s} microseconds after the laser pulse.
\\
We estimate the transmission from the FC positioned in the chamber in front of GANDALPH to the detectors placed after the interaction region to be $\approx$1\%, calculated from the initial intensity of \SI{600}{fA} before the setup and the ion velocity (\SI{135000}{m/s}), derived from $E_{kin}= \frac{1}{2}mv^2$.
This means that there were only 0.1 ions on average in the interaction region. 
Nevertheless, we observed a photodetachment signal as high as \SI{50}{counts/s} of neutralized $^{211}$At in the GANDALPH beam-line when the photon energy was tuned well above the photodetachment threshold. Under these conditions, the combined neutralization and detection efficiency for an ion in the interaction region, which was illuminated by the \SI{10}{kHz} repitition rate pulsed laser light, was \SI{5}{\%}.

\subsection{Accuracy of EA measurements.} The uncertainty in our experiment is dominated by the laser bandwidth of  \SI{12}{GHz}, corresponding to \SI{50}{\micro eV}\cite{Fedosseev_2017}.
In addition, there are several minor effects contributing to the uncertainty: for an \ch{LaB6} surface ionizer, as used in this experiment, the energy spread has been determined to be of the order of  \SI{0.55}{eV}\cite{Kashihira1977SourceIons}. 
This implies a velocity spread of the ions which is compressed due to the acceleration over a high potential in the subsequent ion beam extraction process\cite{Kaufman1976High-resolutionBeams}. %(elaborate?) \\
The compressed velocity spread of the ions is given by the expression  $\Delta v={\Delta W} / {\sqrt{2mW}}$, where $m$ is the ion mass, $\Delta W$ the energy spread of the ions and $W$ the kinetic energy of the ion beam\cite{HANSTORP1995165}.
The velocity spread of the ion beam can be converted to a spread of the frequency of the laser light of $\Delta \nu =  {\Delta v} /{\lambda}$ seen by the ions.   
This results in a frequency Doppler broadening of only a few MHz in the fast ion beam. 
The divergence  of the ion and laser beams and the interaction time will also contribute to the broadening. 
However, this accumulates to uncertainties of less than \SI{10}{MHz}.
Consequently, the uncertainties arising from these minor effects could be ignored and only the laser bandwidth of \SI{12}{GHz} needs to be considered.\\ 
In addition to these statistical errors, some systematic uncertainties arise: 
the Doppler shift due to the velocity difference of ions and photons is very large but it can, as described above, be eliminated to all orders by performing the experiment with both co- and counter-propagating laser and ion beams and calculating the geometrical mean to determine the Doppler-free threshold. Hence, the Doppler shift does not contribute to the uncertainty of the result, barring slight potential angle misalignment of maximum \SI{24}{mrad} as defined by the apertures.
However, uncertainties of the ion beam energy and the wavelength calibration could potentially affect the results.
Such drifts were estimated to be smaller than \SI{20}{\micro eV} by comparing two reference scans on stable $^{127}$I which were performed with the same setup before and after the measurements on astatine.

\subsection{Computational details.}

To achieve an optimal accuracy in the DC-CCSD(T) calculations, all electrons of iodine and astatine were correlated, and all virtual orbitals with energies below \SI{2000}{\text{a.u.}} were included in the virtual space. Fully uncontracted correlation-consistent all-electron relativistic basis sets of Dyall were used\cite{Dya06}. In order to obtain accurate results for the EA, high quality description of the region removed from the nucleus (that will contain the added electron) is important. We have thus augmented the basis sets with two diffuse functions for each symmetry block. Finally, we performed an extrapolation to the complete basis set (CBS) limit, using the scheme of Halkier \textit{et al.}\cite{HalHelJor99} for the DHF values and the CBS(34)\cite{HelKloKoc97} scheme for the correlation contribution.
In the DC-CCSD(T) calculations, the finite size of the nucleus was taken into account and modelled by a Gaussian charge distribution within the DIRAC15 program package\cite{VisDya97}. \\
Full triple and perturbative quadruple (Q) contributions were calculated in a limited correlation space with the valence $6s$ and $6p$ electrons and a virtual orbital energy cutoff of \SI{30}{\text{atomic units}}. It has been previously demonstrated  that higher-order correlation is dominated by the valence contributions\cite{PasEliBor17}, and thus this correlation space is deemed sufficient.
The valence vXz basis sets of Dyall\cite{Dya06} were used, and extrapolated to the CBS limit as above. These calculations were performed using the program package MRCC\cite{MRCC,KalSur01,BomStaKal05,KalGau05,KalGau08} linked to DIRAC15. Full Q contributions evaluated at the v2z level were below \SI{1}{meV} for both systems and were thus omitted.\\ % Breit & QED
Due to the non-instantaneous interaction between particles being limited by the speed of light in the relativistic framework, a correction to the two-electron part of $H_\text{DC}$ is added, in the form of the zero-frequency Breit interaction calculated within the Fock-space coupled cluster approach (DCB-FSCC), using the Tel Aviv atomic computational package \cite{TRAFS-3C}. 
To account for the QED corrections, we applied the model Lamb shift operator (MLSO) of Shabaev and co-workers\cite{ShaTupYer15} to the atomic no-virtual-pair many-body DCB Hamiltonian.
%as implemented into the QEDMOD program. 
This model Hamiltonian uses the Uehling potential and an approximate Wichmann--Kroll term for the vacuum polarization (VP) potential\cite{Blo72} as well as local and non-local operators for the self-energy (SE), the cross terms (SEVP) and the higher-order QED terms\cite{ShaTupYer13}.
The implementation of the MLSO formalism in the Tel Aviv atomic computational package
allows us to obtain the VP and SE contributions beyond the usual mean-field level, 
namely at the DCB-FSCC level.\\
The three remaining sources of error in these calculations are the basis set incompleteness, the neglect of even higher excitations beyond (Q), and the higher-order QED contributions. The first of these is the largest.
We have extrapolated our results to the complete basis set limit, and as the associated error, we take half the difference between the CBS result and the doubly augmented ae4z (d-aug-ae4z) basis set value which is \SI{0.015}{eV}.
We assume that the effect of the higher excitations should not exceed the (Q) contribution of \SI{0.004}{eV}, and that the error due to the incomplete treatment of the QED effects is not larger than the vacuum polarization and the self energy contributions  of \SI{0.003}{eV}.
Combining the above sources of error and assuming them to be independent, the total conservative uncertainty estimate on the calculated EA of At is \SI{0.016}{eV}, dominated by the basis set effects. 

\subsection{Comparison to previous theoretical results.}
Some recent calculations, including our final theoretical value of the EA of At (labelled CBS - DC - CCSDT(Q) + Breit + QED) are compared to the experimental value in Tab. \ref{tableII}. 
Of particular interest is the recent multi-configurational Dirac-Hartree-Fock (MCDHF) study of Si and Fischer\cite{SiFis18}. Including the Breit and the QED corrections and extrapolating systematically in terms of included configurations, they obtained an EA for iodine (\SI{3.0634(24)}{eV}) in excellent agreement with experiment. However, the analogous result for At (\SI{2.3729(46)}{eV}) lies outside the uncertainty of our experiment.
\begin{table*}
\footnotesize
\caption{The electron affinity of astatine from the present calculations in comparison to other theoretical approaches.}
\label{tableII}
\begin{center}
\begin{minipage}{\linewidth}
\centering
\begin{tabular}{ l S l } 
 \hline\hline
 Method & ~~~~~~~~~~~~~EA(At)/eV& Ref. \\ 
 \hline
CBS-DC-CCSDT(Q)+Breit+QED & 2.414(16) & this work \\ 
  MCDHF+SE corr.\footnote{Multiconfigurational Dirac-Fock (MCDF) results corrected using experimental data.} & 2.38(2) & \cite{ChaLiDon10} \\ 
 MCDHF & 2.416 & \cite{LiZhaAnd12} \\ 
DC-CCSD(T)+Breit+QED & 2.412 &   \cite{Borschevsky2015IonizationAt} \\
MCDHF+Extrap.+Breit+QED \footnote{MCDF results extrapolated to complete active space limit} & 2.3729(46) &   \cite{SiFis18}  \\ 
 CBS-DC-CCSD(T)+Gaunt+QECBS-DC-CCSD(T)+Gaunt+QEDD& 2.423(13) & \cite{FinPet19} \\
Experiment & 2.41578(5)  & this work\\
 \hline\hline
\end{tabular}
\end{minipage}
\end{center}

\end{table*}
More recently, another very accurate calculation of the EA of At (and other heavy \textit{p}-block elements) was carried out by Finney and Peterson\cite{FinPet19}, using an approach similar to that employed in this work. 
They obtained an EA of \SI{2.423(13)}{eV}, which is in very good agreement with both the measurement and the prediction of this work. The difference between the two theoretical results is mainly due to the number of correlated electrons (all 85 in the present calculation vs. 25 in Ref.\cite{FinPet19}), the use of the Gaunt correction (instead of Breit) in Ref.\cite{FinPet19} and the lack of the higher excitations in earlier work.

\subsection{Data availability}
The data-sets generated and/or analyzed during the current study are available from the corresponding authors on reasonable request.
\end{methods}

% Bibliography
\subsection{References}
\bibliography{bib}

%% Here is the endmatter stuff: Supplementary Info, etc.
%% Use \item's to separate, default label is "Acknowledgements"

\begin{addendum}
 \item We thank the ISOLDE technical team and the operators for their work converting ISOLDE to a negative ion machine. The Swedish Research Council is acknowledged for financial support. We would also like to thank the Center for Information Technology of the University of Groningen for their support 
and for providing access to the Peregrine high performance computing cluster. N.G and E.R. acknowledge the French National Agency for Research for grants called Programme d$'$Investissements d$'$Avenir (ANR-11-EQPX-0004, ANR-11-LABX-0018). Y.L. acknowledges support from the Office of Nuclear Physics, U.S. Department of Energy under Contract No. DE-AC05-00OR22725.
This project has received funding from the European Union’s Horizon 2020 research and innovation programme under grant agreement No 654002 and by the innovative training network fellowship under grant No 642889. L.F.P is grateful for the support from the Slovak Research and Development Agency (APVV-15-0105) and the Scientific Grant Agency of the Slovak Republic (1/0777/19). R.H. acknowledges support by the Bundesministerium f\"ur Bildung und Forschung (BMBF, Germany) under the consecutive projects 05P12UMCIA, 05P15UMCIA and 05P18UMCIA. 
This work was also supported by the FNPMLS ERC Consolidator Grant no. 64838 and the FWO-Vlaanderen (Belgium) and the GOA 15/010 grant from KU Leuven.  
We would like to acknowledge Kevin Patrice Moles for his help with the design of Fig. \ref{fig:GANDALPH} and \ref{fig:RIB}.
 \item[Author contributions]
V.F., N.G., D.H., E.R., S.R., J.S. conceived the experiment and wrote the proposal; 
L.B., D.H, D.L., A.R.M., S.R., J.S., J.Wa., J.We. designed and constructed GANDALPH;
L.B., D.H., R.H., M.K.K., D.L., A.R.M, S.R., J.S., J.We. setup and operated GANDALPH;
R.A., K.C., D.F., R.G.-R., C.G., R.H., A.K., B.A.M., P.M., M.R., S.R., D.S., M.R.,A.V., S.W. setup and operated  the laser system;
J.B., F.B.P., D.L., J.P.R., S.R. operated the target and ion source;
K.C.,O.F., R.G.-R., D.H., R.H., D.L., Y.L., A.R.M., M.R., R.R., S.R., D.S., J.S., J.Wa., K.W. participated in data taking;
D.H., D.L., S.R., J.S., J.We. analyzed the data;
A.B., N.G., Y.G., E.E., L.F.P., E.R. performed calculations;
N.G., D.H., D.L., B.A.M., U.K., S.R., J.S. wrote the manuscript draft;
A.B., V.F., N.G., D.H., S.R., K.W. coordinated the project and/or supervised the participants;
all authors contributed to the discussion of the manuscript.
 \item[Competing Interests] The authors declare that they have no
competing financial interests.
 \item[Correspondence] Correspondence and requests for materials
should be addressed to D. Leimbach (email: davidleimbach@posteo.de).
\end{addendum}

%%
%% TABLES
%%
%% If there are any tables, put them here.
%%
%\end{multicols}{2}

\end{document}